\documentstyle[11pt,paspconf,epsf]{article}

\begin{document}

\title{An overview of dynamical models for outflows in BALQSOs and Seyferts}
\author{Martijn de Kool}
\affil{JILA, University of Colorado}
\begin{abstract}
In this paper I will review the dynamical models that attempt to 
explain the outflows in QSOs and Seyfert galaxies that are responsible 
for the blue shifted absorption lines observed in some of these objects.

Most models face the difficulty that the absorbing gas appears to
consist of very small, low filling factor clouds, that are likely to
be hydrodynamically unstable, and require an intercloud medium at high
pressure to keep them confined. Two types of intercloud medium have
been proposed:  hot gas and magnetic field. Hot gas confinement leads
to many apparent physical contradictions, and until these are resolved
can not be considered a serious candidate.  Magnetic confinement 
seems more promising, but needs more study to assess its viability.
A third possibility that should be seriously considered in the light of
the confinement problem is that some of the arguments leading to the
conclusion that a small filling factor is required is wrong. As an example,  
Murray et al. showed that a model that allows the ionization
parameter of the absorbing gas to be much higher than generally
thought can avoid the confinement problem.

Three types of acceleration mechanisms have been considered for
 wind-like outflows in AGN: hot gas or cosmic ray pressure 
driven winds, winds
driven by radiation pressure and centrifugally driven magnetic disk winds.
The first two have been shown to be able to
produce absorption lines similar to the observed ones, with 
line radiation pressure requiring the smallest leaps of faith. The
formation of absorption lines in magnetic winds is still unexplored. 
\end{abstract}


\section{Introduction}

The existence of broad absorption line quasars and blue shifted
absorption lines in Seyferts provides clear evidence that high speed 
outflows of gas are common in AGN. The basic questions one would like to
have answered by a theoretical model for these outflows are:
\begin{itemize}
\item What is the origin of the gas flowing out?

\item What is its geometry?

\item What force accelerates it?

\item What determines its ionization state?
\end{itemize}

While answering these questions, a good model should satisfy these
observational constraints:

\begin{itemize}
\item it should produce lines similar to the observed
ones. Significant column densities ( $10^{15}$ to $10^{18}$ cm$^{-2}$)
of both high ionization potential ions ( O VI, N V, CIV etc.) and low
ionization potential ions (O III, Si IV) should be produced.  

\item material has to be accelerated
 up to at least 30,000 km sec$^{-1}$ in QSOs
and 5000 km sec$^{-1}$ in Seyferts

\item the ionization parameter should not vary drastically for
gas with different speeds along the line of sight

\item the outflow should not produce emission at speeds higher than
observed in the emission lines, which presumably implies it has a
covering factor of less than 20 \% (Hamann et al. 1993)

\item the geometry should be such that high speed BAL material occurs outside
the Ly$\alpha$ broad emission line region, since Ly$\alpha$ is seen to be
heavily absorbed by high speed material containing N V ions. 

\item it should explain why there is such a wide range in absorption
line morphology (Turnshek 1988)

\item it should be consistent with the lack of variability in the
structure of the absorption lines in BALQSOs over timescales of 10
years (Barlow 1994)
\end{itemize}

In this paper I will give a critical review of the theoretical models
that have been proposed so far, and discuss to what extent they fulfil
the criteria above.

\section{The state of the absorbing gas and the confinement problem}

The fact that the strongest absorption lines formed in the outflow are
of species like C IV, O VI, O III, Si IV and N V suggest that if the
absorbing gas is irradiated by a ``standard'' AGN spectrum (e.g. Mathews and
Ferland 1987) the
ionization parameter $U$ (density of photons below the Lyman limit
divided by the hydrogen density) is of the order 
$ U \sim 0.1$. This leads immediately to an estimate of the density of
the absorbing gas of
\begin{equation}
n_{abs} \sim 10^{10} L_{46} R_{17}^{-2} U_{-1}\ {\rm cm}^{-3}
\end{equation}
with $ L_{46}$ the luminosity of the central engine in units of
$10^{46}$\ ergs/sec, $ R_{17}$ the size of the region in which the
absorption lines are formed in units of $10^{17}$\ cm, 
and $U_{-1}$ the ionization parameter normalized to 0.1.
Estimates of the total column density of the absorbing gas are
difficult to obtain due to saturation of the lines, but are typically in
the range $N_{abs} \sim 10^{21} - 10^{23}  {\rm cm}$, so that the
total thickness of the absorbing layer is
\begin{equation} 
D_{abs} \sim {{N_{abs}}\over{n_{abs}}} \sim 10^{13} N_{23} L_{46}^{-1}
R_{17}^{2} U_{-1}\    {\rm cm}\ .
\end{equation}
From the fact that both the continuum and the broad emission lines are
seen to be absorbed, we know that the absorbing region lies outside or
is cospatial with the BEL region, we have 
$R_{abs} \sim R_{BEL} \sim 10^{17}{\rm cm}$. 
Thus, comparing the size of the absorbing
region with the thickness of the absorbing layer we arrive at a very
small filling factor for the absorbing gas, 
\begin{equation}
f_{abs} \sim {{D_{abs}}\over{R_{abs}}} \sim 10^{-4} N_{23} L_{46}^{-1} R_{17} U_{-1} 
\end{equation}
If we require additionally that the flow gives rise to the smooth
absorption troughs observed in some BALQSOs that extend over 
hundreds to thousands of thermal
line widths, it is clear that we will need hundreds to thousands of
clouds in our line of sight, implying very small sizes 
($\sim 10^{10}{\rm cm}$) for individual clouds. 

What are the consequences of this low filling factor? It is
clear that there has to be a confining medium around the clouds to
keep them from expanding and becoming too highly ionized, since the
sound crossing time of the clouds is very short. There are two obvious
choices for this confining medium: hot gas or magnetic field.

\subsection{Hot gas confinement}

Confinement by hot gas would be the simplest explanation, but it leads
to a large number of problems( Weymann et al. 1985, Begelman et al. 1991). 
If there is any
velocity difference in the outflow between the clouds and the hot gas,
the timescale for destruction of the clouds by hydrodynamical
instabilities such as Kelvin-Helmholtz or Rayleigh-Taylor is very
short compared to the acceleration time. 
The flow could consist of a true
two-phase medium in which cool clouds could continuously reform, but
unless the densities are higher than we infer, 
the timescale to cool down from the hot phase
is longer than the crossing time of the absorbing region.

Because of the very low filling factor of the cool clouds, the mass of
gas in the hot phase will be larger than in the clouds 
unless it has a very high temperature, $\sim 10^9 K$. 
If the temperature of the hot phase would be around the Compton
temperature of the AGN radiation field, the mass loss in the hot phase
would be uncomfortably large, of the order of tens of solar masses per
year. However, gas at a temperature of $\sim 10^9 K$
is not able to confine the small clouds thought to exist in the
absorbing region, because the mean free path of the hot particles is
longer than the size of the clouds.

Finally, there is a problem with the fact that the ionization
parameter of the absorbing clouds is observed to remain
 approximately constant as they are accelerated (e.g.Turnshek 1988). 
Any hot gas flowing out with the clouds
will cool down quickly due to adiabatic expansion losses, causing the
confining pressure to drop much too rapidly to keep $U$ approximately
constant. Thus, any hot gas confinement model has to have some way to
release a lot of energy into the hot gas as it is being accelerated.

\subsection{Magnetic confinement}

Confinement by magnetic field avoids several of the problems
associated with hot gas. If the clouds travel along magnetic field
lines, hydromagnetic instabilities are likely to be much less severe,
and the drag forces could be very small, although these effects have
not been studied in much detail yet. A magnetic field
could provide confinement in the direction perpendicular to the field,
but other processes have to be invoked for confinement along the field. 
One advantage is that the cloud can expand only in one dimension, so that
it will take a much longer time for the density to drop than in the
case of spherical expansion. Confinement along the field could
possibly be provided by ram pressure or radiation pressure from the
central source, since the development of the Rayleigh-Taylor
instability normally associated with this process will be suppressed
by the magnetic field. There is in fact some indication that the
clouds are being pushed, since the small size of the clouds is consistent
with the scaleheight in the cool gas assuming that they experience an
``artificial gravity'' equal  to their mean acceleration 
$v_{BAL}^2 / R_{BAL}$ (de Kool and Begelman 1995). 
Another possibility is that the flow is
self-confining in the direction along the flow, i.e. it consists of very
thin flow tubes with a very low filling factor and small size perpendicular
to the field, that are completely filled with cool material. Such a
picture would fulfil all the constraints derived above.

Simple magnetic field models (such as completely radial or azimuthal
field lines) also can not reproduce a confining pressure that is
approximately constant as the clouds are accelerated, since the field
strength drops too rapidly. More complicated field geometries are
clearly required.

\section{Acceleration Mechanisms}

\subsection{Pressure driven winds}

The large drag forces that the absorbing clouds experience when they
move relative to any confining gas led Weymann et al. 1985 to the
conclusion that the clouds are likely to be dragged along in an
accelerating confining medium rather than being accelerated
themselves. The low density and high pressure in the confining gas
make it a good candidate for a pressure driven wind. 
Pressure driven winds typically reach end speeds of a few times
the sound speed in the wind, so the high velocity observed in BALQSO
outflows requires either an extremely hot wind ($T > 10^9 K$), 
or a wind driven by non-thermal particles such as cosmic rays.

Two models have been proposed along these lines. The first was by Weymann
et al. 1982, which attempted to explain the broad emission line region
 as a hot outflow with
embedded clouds, driven by relativistic electrons diffusing out of the
central engine. The second model (Begelman et al. 1991) was
specifically aimed at explaining the broad absorption lines. Here the
primary cosmic rays were highly relativistic neutrons that are
produced in the central engine, that can travel out to parsec scale
before they decay to protons and couple locally to the magnetic field.
Because the neutrons decay over a wide range of radii, there is a
steady energy injection into the wind as it is being accelerated,
compensating for the adiabatic losses that would otherwise make the 
confining pressure drop too fast.
This model can produce BALs with characteristics similar to the
observed ones, but it does require significant fine tuning of the
parameters. Note that cosmic rays have very long mean free paths, and
will penetrate the BAL clouds easily, so that the cosmic ray pressure 
itself can not provide confinement. Thus, dissipation of cosmic ray energy
into the thermal intercloud gas (to equipartition) seems to be required anyway.  

The origin of the clouds in these models is not clear. For stability
reasons, it would be preferable if large velocity differences between
the clouds and the confining medium would never develop, and from this
point of view it is most attractive to inject the clouds at the base
of the wind and let the ensemble accelerate together. On the other
hand, there are many BAL profiles in which the absorption sets in
sharply at some high velocity, which would be hard to understand in a
radial outflow picture. An alternative injection mechanism is that
the wind is accelerated on smaller scales, and hits some large
(molecular?) cloud at high speed, ripping off material from the
surface. In this picture, we would only see a BAL if we are looking
through the wake of such a cloud. It is doubtful however that the
clouds of cold material could survive long enough to be accelerated
to velocities as high as observed, because (at least simple) estimates
of the hydrodynamical destruction time are much shorter than the
acceleration time. Numerical models along these lines
(Schiano et al. 1995) have not been conclusive.

Observationally, we do not have much direct evidence for the existence
of a very hot wind in BALQSOs. Stocke et al. (1992) have argued that the 
radio emission from radioquiet BALQSOs and Seyferts is consistent with a hot
wind model, and exhibits the proper scaling of outflow speed with luminosity.

\subsection{Centrifugally driven magnetic disk winds}

The studies of this outflow mechanism have so far been restricted to the
formation of broad emission lines (Emmering et al. 1992, Bottorff et
al. 1997) or dusty winds at large disk radii that could form the
obscuring torus in Seyfert 2 galaxies (K\"onigl and Kartje 1994).
Clearly it is also a good candidate for the origin of
blueshifted absorption lines, but there are as yet no models exploring the
formation of absorption lines in such flows in detail.

All these models are based on the basic self-similar outflow model
proposed by Blandford and Payne (1982). It should be kept in mind that
this model is highly simplified to make it mathematically tractable. 
There are no reasons why the flow should be self-similar,
since self-similarity is only obtained for a special set of boundary
conditions that is not preferred over others. 

In this model (see Figure 1)
cool clouds of gas emerge from the surface of the disk
threaded by magnetic field lines that are anchored in disk. If the
angle between the field line and the disk is small enough, the cloud
will move outward but is forced to corotate with the point in the disk
where the field line is anchored, making it rotate faster than local
Kepler speed. This causes a centrifugal force on the cloud, that tends
to move it to even larger radii. The speeds attained scale with the
Kepler speed at the field line footpoint, and with the ratio of
magnetic flux to mass flux on the surface of the disk. For reasonable
values, outflow speeds of the order 3 times the Kepler speed are
reached.

\begin{figure}[tp]
\mbox{}\hfill  \epsfxsize=10cm\epsfbox{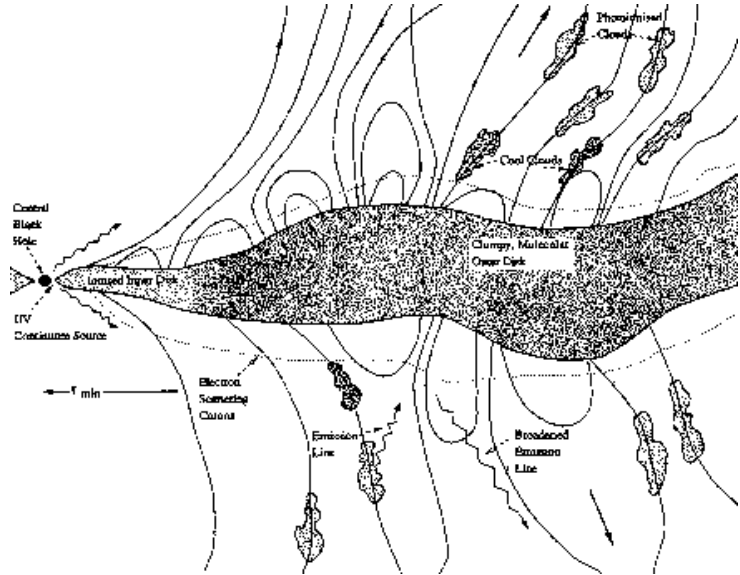} \hfill\mbox{}
\caption[$EBS schema]{ A schematic representation of magnetic disk
 wind model of
 Emmering et al. (1992).}
\label{EBS}
\end{figure}

Even the simple self-similar model has about ten free
parameters to specify the exact flow geometry (Bottorff et al. 1997).
In the light of this, it may not be surprising that this model is able
to fit the BEL profiles of many sources quite well. Emmering et
al. 1992 did however need to invoke an extra source of scattering for
the line photons in order to obtain wide enough line profiles if the
base of the outflow is restricted to two decades in radius on the
surface of the disk.  
An independent test of this model was performed by Bottorff et al. who 
showed that for NGC 5548, a
model of this kind for which the parameters have been fixed by
fitting the line profile alone can reproduce the response of the line profile
to continuum variability quite accurately (see these proceedings).

To what extent this type of outflow can also be responsible for BALs is
not clear. There may be a problem in obtaining high enough speeds,
since BAL speeds are typically several times higher than BEL speeds,
and the model already has problems obtaining BELs as broad as
observed. Also the basic geometry of the self-similar model is not
favorable to explain the fact that high velocity BAL material is
obscuring low velocity material, since in these models the high
velocities occur on much smaller scale than the low velocities,
typically scaling with $v \propto r^{-{1 \over 2}}$. Thus the high velocity
material with BAL-like speeds would typically be expected to be 10-100
times closer in than the material with typical broad emission line speeds.
For Seyferts however, which show absorption lines that typically
have speeds similar to the broad emission line width, this seems a
more promising model.
A theoretical problem may be the neglect of the effects of
radiation pressure on the clouds, which should be important
unless the clouds are very optically thick.

\subsection{Outflows driven by radiation pressure on dust}

One suggestion about the nature of BALQSOs is that they are QSOs
near the end of their forming stage, that are blowing out the remaining
circumnuclear gas after the central engine has turned on (e.g. Voit et
al. 1993). 
An important driving force for this outflow would be radiation pressure on
dust. This model is not very popular, because the
argument that the upper limit on the covering factor of the BAL region
is about equal to the observed fraction of BALQSOs to non-BALQSOs
suggests that all QSOs must have BAL regions. However, the point about
the likely importance of radiation pressure on dust is a very good one. 

Scoville and Norman (1995, see also these proceedings) 
have worked out a model in which BAL material
derives from the stellar winds of cool giants, that is 
accelerated outward by radiation pressure acting on the dust in the
wind. The high outflow speeds of $\sim$ 30,000 km sec$^{-1}$ are a natural
consequence of this model, given that the dust opacity is relatively well
known, and the starting radius of the outflow must be approximately
equal to the radius ($\sim$\ 1 pc) where the equilibrium temperature
of the dust is equal to the evaporation temperature of $\sim$ 1800
K. A weak point in this model appears to be the confinement of the
cool gas in the wind trails from the giants, which is due to
the trails being compressed between the radiation pressure from the
inside, and the ram pressure due to the non-comoving 
surrounding medium from the
outside, which would appear to be very unstable. From an observational
point of view, there is the problem that many high-ionization BALQSOs
show no signs of dust absorption in their UV spectra 
(e.g. Weymann et al. 1991, Korista et al. 1992).

\subsection{Line radiation pressure driven winds}

There are several compelling arguments that acceleration by UV line
photons plays an important role in BALQSOs. It has the advantage over
the other models that the source of the acceleration is directly
observed. Making a simple estimate of the total momentum in the
outflow:
\begin{equation}
\dot M v_{wind} \sim 10^{33} f_{-1} N_{23} R_{17} v_9^2 \ {\rm g\ cm\ s^{-2}} 
\end{equation}
with $f_{-1}$ the covering fraction of the BAL outflow in units of 0.1,
$N_{23}$ the column density of the BAL region in units of 
$10^{23}{\rm cm{-2}}$, $R_{17}$ its size in units of $10^{17}$\ cm ,  and
$v_9$ the wind speed in units of $10^4$ $\rm km s^{-1}$, and comparing
this to the momentum absorbed from the radiation field in the BAL troughs:
\begin{equation}
{{\Delta L}\over c} \sim {{f_{cov} N_{line} {{W_\lambda}\over \lambda} L}\over c}\sim 10^{34} \dot L_{46} f_{-1} N_{1} v_9\ {\rm g\ cm\ s^{-2}}
\end{equation}
where $ N_{1}$ is the number of important driving lines divided by 10,
and ${{W_\lambda}\over \lambda} \sim {v_{wind} \over c}$,
we find that line radiation pressure must play an important
role. Furthermore, there is some direct evidence for the dynamical
importance of line radiation by the occurrence of line locking in some
Si IV profiles (Turnshek et al. 1988) and extra acceleration of the
flow at velocities where the Ly$\alpha$ broad emission line overlaps
with the N V absorption, resulting in a steeper velocity gradient and
a reduced optical depth in the BAL profiles of other species (Arav and
Begelman 1994). The standard problems associated with small
clouds and low filling factors (drag, hydrodynamical
destruction, confinement with constant U) also plague this model, but 
recent studies of the instabilities occurring in line driven winds
from O-stars may provide some solutions to these problems (e.g. Owocki
1994, and the contribution by Feldmeier and Norman in these proceedings).
 
\subsubsection{Early models}

Because of the similarity of many BAL profiles with stellar P-Cygni
profiles of hot stars with fast winds driven by line radiation, this
model was in fact among the first to be considered for BALQSOs. In
analogy with hot stars, Drew and Boksenberg (1984) constructed
spherically symmetric wind models with a filling factor of 1. They
were not very successful in reproducing the observed profiles because of
the very reasons that led to the constraints on the filling factor
above: unrealistically high mass loss rates are necessary to keep the
density high enough, and thus the ionization parameter sufficiently low to avoid
heating the gas to too high temperatures, and it was very difficult to
keep the ionization parameter approximately constant as the flow is
accelerated. Also the model could not account for the absence of
emission at high velocities, a consequence of the spherical symmetry
that was assumed.

An early radiatively driven disk wind model for the broad emission
line region was proposed by Shlosman et al. (1985). In this model the
wind is not driven radially outward, but vertically up from the
UV-emitting part of the accretion disk. While it is still in the
shadow of the atmosphere of the inner disk the wind is in a single,
cool phase. As it gets exposed to the full hard AGN continuum above
the atmosphere it becomes thermally unstable and
 turns into a two-phase state, with cool clouds embedded in a hot gas.
This effectively launches clouds with a velocity perpendicular to the
disk, and it is assumed that after this radiation pressure is no longer
important. In this model, we would expect to see broad absorption
lines under most viewing angles, with the highest velocities
occurring when we are observing the disk face-on, in contrast to most
other disk wind models where the wind is visible only if our line of
sight is close to the disk plane.  

\subsubsection{Line driven wind models with low filling factor}

More recently, Arav et al. (1994) reinvestigated the possibility of
line radiation pressure driven winds in BALQSOs. They decided to
neglect all the problems associated with low filling factors, and just
{\it assumed} that the absorbing clouds are infinitely small, and
embedded in a massless confining medium that exerts no drag forces,
and has the correct run of confining pressure with radius to be
consistent with observations. Once these assumptions are made,
the classic CAK theory for line driven winds (Castor et al. 1975) can
be applied to derive the properties of the wind. This theory is
remarkably successful and correctly predicts the important driving
lines, and the final velocity if it is assumed that the flow starts at
radii comparable to the broad emission line region, and is exposed to the
optical to X-ray continuum produced close to the center of an AGN.
They also pointed out that if the confining pressure varies with
radius in such a way that the ionization parameter is approximately
constant, the gradient of this pressure should play a non-negligible
role in the acceleration, and a model taking this into account was
presented.

A possible candidate for the massless confining medium with just the
right properties was suggested by de Kool and Begelman (1995). Their
model is a hybrid between the radiation pressure and magnetic disk wind
models, that is based on the simple estimate that line radiation pressure will
accelerate clouds much faster than the magnetic centrifugal effect (if
they are not too optically thick). Just as in the model of Emmering et
al. (1992) cool matter emerges from the disk threaded by magnetic
field. As it is exposed to the central continuum, the matter is heated and
expands until it is in pressure equilibrium with the magnetic field
around it. Line radiation pressure then accelerates the clouds,
dragging the magnetic field along with it (see Figure 2). 
The radiation pressure alone would accelerate the flow in a very thin 
wedge over the surface of the disk. However, all of the magnetic flux emerging
from the inner parts of the disk is forced into this wedge, and if it
becomes too thin large magnetic pressure gradients develop that will
widen the outflow wedge again. Thus an equilibrium configuration will
develop in which the magnetic pressure gradients balance the
radiation pressure, which is only possible if the magnetic pressure is
a significant fraction of the radiation pressure. This is just what is
required to obtain an ionization parameter in the clouds that is 
self-regulating and approximately constant.

\begin{figure}[tp]
\mbox{}\hfill  \epsfxsize=10cm\epsfbox{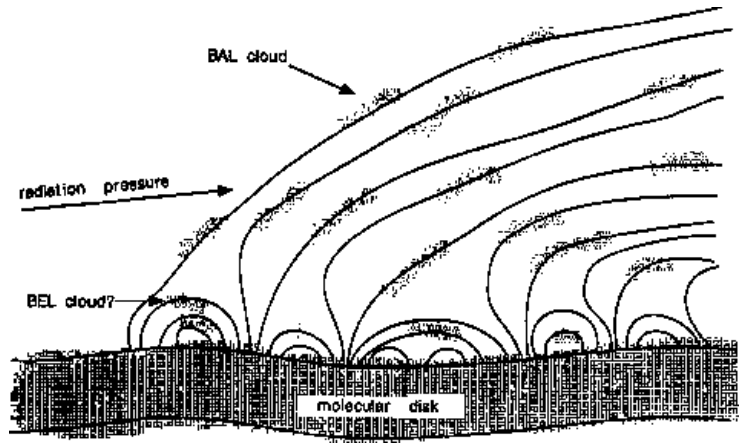} \hfill\mbox{}
\caption[dKB schema]{The radiatively accelerated and magnetically confined 
disk wind model of de Kool and Begelman (1995). The radiation from
the central source compresses the magnetic field until its pressure is
comparable to the radiation pressure, thus providing a roughly
constant ionization parameter.}
\label{dKB}
\end{figure}

\subsubsection{The disk wind model of Murray et al.}

The most recent example of a radiatively driven disk wind model is the
one by Murray et al. (1995). This differs radically from all the other
ones, since it assumes that the filling factor of the absorbing
material is unity. In order to avoid the earlier arguments against
this, the authors argue that the ionization parameter of the BAL gas
is about three orders of magnitude higher than previously
assumed. This is possible if one assumes that the ionizing spectrum
irradiating the BAL gas is not the the full central continuum,
but rather one that has already passed through a high column density
highly ionized absorber (Figure 3). This would absorb the far UV 
and the soft X-rays out of the spectrum, and there would
not be enough hard photons left to ionize  the observed BAL species
like C IV completely away. A very strong point of this model is that
indeed all BALQSOs appear to have strong absorption in their soft
X-ray spectra, since ROSAT observations have shown them to be very
weak soft X-ray emitters (Green and Mathur 1996). 
In the one case where a column density could be determined (PHL 5200,
Mathur et al. 1995) a value consistent with the ones needed to
keep the ionization low was found. In subsequent papers (Chiang and Murray
1996, Murray and Chiang 1997) it is argued that the emission lines 
formed at the base of the disk wind can also account for the 
reverberation mapping results and broad emission line profiles from
Seyfert galaxies (see also these proceedings).  

\begin{figure}[tp]
\mbox{}\hfill  \epsfxsize=10cm\epsfbox{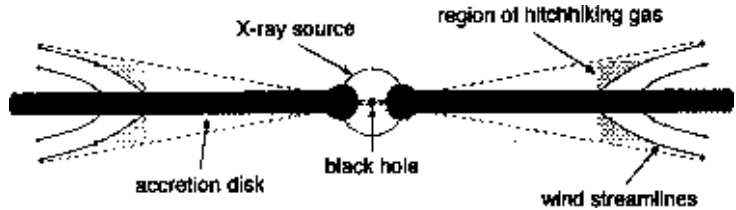} \hfill\mbox{}
\caption[Murrays schema]{ A schematic representation of the disk wind
model by Murray et al. (1995). The material labelled ``hitchhiking
gas'' is what screens the disk wind from the hard X-rays from the  
central source}
\label{Murray}
\end{figure}

This is a very attractive picture because it avoids all the problems
associated with clouds. However, a few problems with this model have
not been satisfactorily resolved. The first one is that the model
requires that the BALs are formed at very small radii $\sim 10^{16}$\
cm. The reason for this is that the highly ionized wind is not as
easily accelerated because of the low abundance of the ions that drive
the wind. To obtain speeds as high as observed in BALQSOs, it is
necessary to start the flow very close in. Since the evidence that the
BEL region lies inside or is cospatial with the BAL region is
incontrovertible, this also implies that the BEL region in QSOs
has a much much smaller size than commonly accepted. With these small
scales, the crossing time of the BAL region is only a month or so, and it
is difficult to understand that the highly complex kinematical
structures that BALs often exhibit do not appear to vary on timescales
of 10 years (Barlow 1994). Even if the absorption components we see
are not actual cloud complexes being accelerated
 but rather standing flow patterns, it
would be likely that they are anchored to certain positions on the
disk, and the short rotation period of the disk would be likely to
cause changes on the observed timescale.
Any structure in the wind that is due to the line driving instability
is not likely to be axisymmetric, and the analogy with O-star winds suggest
that such structures would be seen to move in a few dynamical times (Owocki
1994, Braun and Milgrom 1990).

It is not clear if the full range of ions observed to show
BAL profiles can be explained by the very high ionization parameters
required by this model. As an example, the ionization model presented 
by Murray et al. would have difficulties explaining the
presence of broad absorption lines from ions with ionization
potentials like O III or lower. It would also be interesting to explore
the consequences of the small broad emission line region in more
detail, to see if the proper line ratios can be obtained, and if the
density limits obtained from semi-forbidden lines are not violated.

The origin and dynamics of the highly ionized
absorber assumed to exist between the hard continuum source and the
wind needs to be worked out in more detail. 
Murray et al. argue that it is a ``failed wind''
which attempts to start from the disk at radii interior to the real
wind, which becomes too highly ionized to achieve escape speed but
which could still get far enough above the disk surface to shield the
outer wind. Their derivation of the dynamics is  very
approximative, and a more rigorous calculation (which would involve some
fairly complicated 2-D radiation hydrodynamics) would lend more
credibility to this model.

\section{Conclusions}

It is clear that the problem of the origin of the blueshifted UV
absorption lines in AGN is far from being solved. Looking back at the
the basic questions posed in the introduction, it appears that there
are clear favorites for the accelerating force (
radiative acceleration by UV line photons) and geometry (disk-like,
compare the great similarities between figures 1, 2 and 3), but the
origin of the BAL material and the regulation mechanism for the
ionization parameter are still very uncertain.
One of the outstanding problems is the lack
of variability in the kinematic structures in the BALQSO line
profiles, which is beginning to be an embarrassment even for the
models operating on parsec scale, since small velocity shifts in
the components could have been easily detected. No model has as yet
attempted to explain the highly irregular structure of the absorption
line profiles, with both very smooth broad and some quite
narrow ($\sim$ a few hundred km sec$^{-1}$) components.

It seems likely that most of the processes that play a role in the
formation of the absorption lines have been recognized. It now remains
to look at them in much more detail to see if the primitive models
that have been proposed so far can really fit all the observational
constraints.

\acknowledgments
I thank Nahum Arav and Mitch Begelman for some useful discussions.
This work was supported by NSF grant AST-9528256 and NASA grant NAGW-3838.

\begin{question}{Brad Peterson}
Suppose that broad emission line clouds are not confined and can
freely expand (e.g. expanding atmospheres of stars). The clouds,
driven outward by radiation pressure, expand and become optically
thin, producing the optically thin component of the broad-line
emission. These clouds will continue to expand, and will heat up to
the Compton temperature (higher in Seyferts  than in QSOs because of
the flatter $\alpha_{o-x}$); in Seyferts, you get a highly ionized
warm absorber, and in QSOs you get high velocities, but lower
ionization levels like BAL clouds. What are the major failings of this
very simple model?

\end{question}

\begin{answer}{Martijn de Kool}
I think that the Compton temperature in QSOs would still be much
too high to be consistent with the ionization state of the BAL gas.
\end{answer}
\end{document}